\documentclass[
3p,twocolumn,
amsmath,
amssymb
]{elsarticle}

\usepackage{graphicx}% Include figure files
\usepackage{dcolumn}% Align table columns on decimal point
\usepackage{bm}% bold math
\usepackage{subfigure}
\usepackage{hyperref}
\usepackage{amsmath}
\usepackage{hyphenat}
\usepackage{amsfonts}
\usepackage{booktabs}
\usepackage{amssymb}
\usepackage[usenames]{color}
\usepackage{float}
\usepackage{mathrsfs}
\usepackage{tabularx}
\usepackage{midfloat}
\usepackage{lineno}

\newcommand{\lan}{\langle}
\newcommand{\ran}{\rangle}
\newcommand{\be}{\begin{equation}}
\newcommand{\ee}{\end{equation}}
\newcommand{\bea}{\begin{eqnarray}}
\newcommand{\eea}{\end{eqnarray}}

\hypersetup{bookmarks=true, colorlinks=true,linktoc=page, linkcolor=blue, citecolor=blue, urlcolor=blue }

\journal{Physics Letters B}

\begin{document}

\begin{frontmatter}

\title{Study of $\Delta$ excitations in medium-mass nuclei with peripheral heavy ion charge-exchange reactions}

%\title{Elsevier \LaTeX\ template\tnoteref{mytitlenote}}

\author[1,2,3]{J. L. Rodr\'{i}guez-S\'{a}nchez \corref{mycorrespondingauthor}}
\cortext[mycorrespondingauthor]{Corresponding author: joseluis.rodriguez.sanchez@usc.es}
\author[1,2]{J.~Benlliure}
\author[4]{I.~Vida\~{n}a}
\author[5]{H.~Lenske}
\author[3]{C.~Scheidenberger}
\author[1]{J.~Vargas\fnref{fn0}}
\author[1,2]{H.~Alvarez-Pol}
\author[3]{J.~Atkinson}
\author[3,6]{T.~Aumann}
\author[1]{Y.~Ayyad\fnref{fn1}}
\author[1]{S.~Beceiro-Novo\fnref{fn3}}
\author[3]{K.~Boretzky}
\author[1,2]{M.~Caama\~{n}o}
\author[7]{E.~Casarejos}
\author[1,2]{D.~Cortina-Gil}
\author[1]{P.~D\'{i}az Fern\'{a}ndez}
\author[3,8]{A.~Estrade\fnref{fn4}}
\author[3]{H.~Geissel}
\author[3]{E.~Haettner}
\author[3]{A.~Keli\'{c}-Heil}
\author[3]{Yu.~A.~Litvinov}
\author[1]{C.~Paradela\fnref{fn2}}
\author[1]{D.~P\'{e}rez-Loureiro}
\author[3]{S.~Pietri}
\author[3]{A.~Prochazka}
\author[3]{M.~Takechi\fnref{fn5}}
\author[3,9]{Y.~K.~Tanaka}
\author[3]{H.~Weick}
\author[3]{J.~S.~Winfield}

\address[1]{Universidad de Santiago de Compostela, E-15782 Santiago de Compostela, Spain}
\address[2]{Instituto Galego de F\'{i}sica de Altas Enerx\'{i}as, University of Santiago de Compostela, E-15782 Santiago de Compostela, Spain}
\address[3]{GSI-Helmholtzzentrum f\"{u}r Schwerionenforschung GmbH, D-64291 Darmstadt, Germany}
\address[4]{INFN, Sezione di Catania, Dipartimento di Fisica ``Ettore Majorana", Universit\`a di Catania, I-95123 Catania, Italy}
\address[5]{Institut f\"{u}r Theoretische Physik der Justus-Liebig Universit\"{a}t Giessen, D-35392 Giessen, Germany}
\address[6]{Institut f\"{u}r Kernphysik, Technische Universit\"{a}t Darmstadt, D-64289 Darmstadt, Germany}
\address[7]{Universidad de Vigo, E-36200 Vigo, Spain}
\address[8]{Saint Mary's University, Halifax, Nova Scotia B3H 3C3, Canada}
\address[9]{High Energy Nuclear Physics Laboratory, RIKEN, Saitama 351-0198, Japan}

\fntext[fn0]{Present address: Universisdad Santo Tom\'{a}s, 15001 Tunja, Colombia}
\fntext[fn1]{Present address: National Superconducting Cyclotron Laboratory, Michigan State University, East Lansing, MI 48824-1321, USA}
\fntext[fn2]{Present address: EC-JRC, Institute for Reference Materials and Measurements, B-2440 Geel, Belgium}
\fntext[fn3]{Present address: Department of Physics, Michigan State University, East Lansing, MI 48824-1321, USA}
\fntext[fn4]{Present address: Department of Physics, Central Michigan University, Mount Pleasant, MI 48858, USA}
\fntext[fn5]{Present address: Niigata University, 950-2181 Niigata, Japan}

\begin{abstract}
Isobaric single charge-exchange reactions, changing nuclear charges by one unit but leaving the mass partitions unaffected, 
have been for the first time investigated by peripheral collisions of $^{112}$Sn ions accelerated up to 1\textit{A} GeV at the GSI facilities. 
The high-resolving power of the FRS spectrometer allows us to obtain $(p, n)$-type isobaric charge-exchange cross sections with an uncertainty
of $3.5\%$ and to separate quasi-elastic and inelastic components in the missing-energy spectra of the ejectiles. The inelastic component is associated
to the excitation of the $\Delta$(1232) isobar resonance and the emission of pions in s-wave both in the target and projectile nucleus, 
while the quasi-elastic contribution is associated to the nuclear spin-isospin response of nucleon-hole excitations.
An apparent shift of the $\Delta$-resonance peak of $\sim$63 MeV is observed when comparing the missing-energy spectra obtained from the measurements with proton and carbon targets. A detailed analysis,
performed with a theoretical model for the reactions, indicates that this observation can be simply interpreted as a change in the relative magnitude between the contribution of the excitation of the resonance
in the target and in the projectile.
\end{abstract}

\begin{keyword}
\texttt Baryon resonances  \sep inverse kinematics \sep fragment separator FRS
\end{keyword}

\end{frontmatter}
%\linenumbers

%%%%%%%%%%%%%%%%%%%%%%%%%%%%%%%%%%%%%%%%%%%%%%%%%%%%%%%
\section{Introduction}

In the past, many experimental and theoretical attempts have been made to clarify 
the complexities of the interplay of nucleons and excitations of the nucleon in a nuclear many-body environment.
Although the former experimental and theoretical studies especially at the SATURNE/DIOGENE facility at Saclay and at the Synchro-Phasetron at Dubna have collected a multitude of valuable data,
the issue is far from being resolved. Moreover, in-medium resonance physics is of central importance to understand the meson yields with and without strangeness observed in central heavy ion collisions,
e.g. measured at the (former) FOPI and HADES experiments at the GSI facility. The status and achievements of the past work in elementary 
meson production reactions on the nucleon to excite resonances in elementary reactions 
on the nucleon and in ion-ion collisions were reviewed recently in \cite{Lenske:2018bgr}. 
A review on the status and prospects of heavy-ion single and double charge exchange reactions as tools for nuclear spectroscopy can be found in Ref.~\cite{Lenske:2019iwu}.

Central collision experiments probe nuclear matter at high density and finite temperature which is an advantage for obtaining high multiplicities.
For detailed spectroscopic studies, however, reactions under less violent conditions are of advantage. Peripheral heavy ion reactions, as those reported in this letter,
are providing the proper conditions for such investigations. By a proper choice of projectile and target, of incident energy together with a suitable selection of reaction products,
peripheral reactions can be designed to investigate only a few specifically selected degrees excited in the interacting nuclei. 
Thus, peripheral heavy ion reactions provide conditions which, in principle, allow to address questions on the in-medium properties 
of excited states of the nucleon such as the long studied $\Delta$-isobars. In addition, the study of the $\Delta$-isobars are crucial for the understanding of three body forces \cite{krebs}.
To the best of our knowledge, nucleon resonances may also play a major role for
the unsolved issue of the quenching of Gamow-Teller strength \cite{Ericson73,Sakai99,liang}. 
More recently, it was realized that excited baryons should be taken into account in $\beta$-equilibrated neutron star matter,
possibly changing the composition of the matter in the deeper layers of a neutron star \cite{feng2016,drago14,drago16,cai15}.

One of the most important issues in the study of intermediate energy charge-exchange reactions is the downward energy shift of the $\Delta$-resonance peak
position by $\sim$70 MeV in nuclear targets as compared to proton targets. 
The physical significance of this shift was first pointed out by Contardo \textit{et al.} \cite{cotardo86} in their investigations of the $(^{3}He,t)$ reaction at the Laboratoire
National SATURNE in Saclay (France). However, this phenomenon also persists in other charge-exchange collisions,
such as the $(p, n)$ reaction at kinetic energies of around 800~MeV \cite{Bonner78,Gaarde91}, and in $\pi$-nucleus reactions \cite{Oset82}.
This energy shift does not have a single cause but is due to the combination of several effects that include: the Fermi motion of the nucleons
and $\Delta$-isobars in the nuclear medium \cite{Guet89,Jain90}, $\Delta$ conversion processes through $\Delta$N interactions \cite{Chumillas07}, and the excitation of the $\Delta$
in the projectile \cite{Oset89}. By combining nuclear structure and reaction studies, 
the first quantified microscopic description of the SATURNE data was provided by Osterfeld and Udagawa \cite{Osterfeld:1991ii,Udagawa:1994ym}, 
who found that ion-ion initial and final state interactions and residual nucleon hole-$\Delta$ particle interactions
could be the major sources for the observed apparent mass shift. 
Hence, the interpretation of data requires a sound theoretical approach, accounting equally well for reaction dynamics and nuclear structure effects.

In this letter, we use isobaric charge-exchange reactions induced by relativistic beams of stable medium-mass projectiles
as a tool to study the excitation of nucleon resonances in nuclear matter.
These reactions where the projectile and target nuclei exchange one nucleon and the projectile and ejectile contain the same number of nucleons 
guarantee that, with a large probability, the charge-exchange process is mediated by a quasi-free nucleon-nucleon collision~\cite{bachelier86,chiba91}. In the case of inelastic collisions
the excited nucleon resonances, produced in the overlapping region between projectile and target nuclei, decay by pion
emission. To preserve the initial number of nucleons in the projectile remnant, pion decay should not induce any sizable
excitation energy in the projectile.

The aim of this letter is twofold. First, we present the experiment that was performed at GSI using beams of $^{112}$Sn to
induce ($p, n$)-type isobaric charge-exchange reactions. The ejectile fragments are identified unambiguously in atomic and mass number
on an event-by-event basis by using the zero-degree magnetic spectrometer fragment separator FRS~\cite{Geissel1992}.
This spectrometer also allows us to measure
with high accuracy the momentum of the ejectiles. Here, we will detail the data analysis and show the first results.
Secondly, we compare our experimental data to some 
sophisticated model calculations in order to understand the energy shift observed in the $\Delta$-peak.

%%%%%%%%%%%%%%%%%%%%%%%%%%%%%%%%%%%%%%%%%%%%%%%%%%%%%%%
\section{Experimental approach}
\label{sec:exp}

\begin{center}
\begin{figure}[t]
\begin{center}
\includegraphics[width=0.47\textwidth,keepaspectratio]{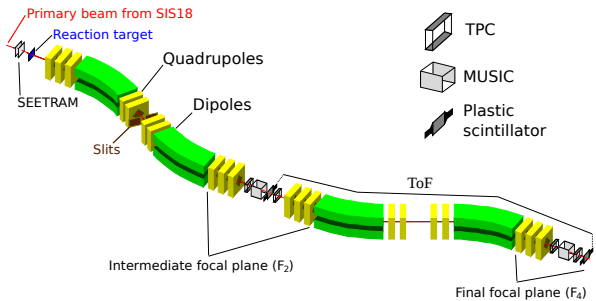}
\caption{(Color online) Schematic view of the FRS experimental setup used in the present work with the reaction target at the entrance of the FRS spectrometer
to induce the $(p, n)$ isobaric charge-exchange reactions.
}
\label{fig:1}
\end{center}
\end{figure}
\end{center}

The experiment was performed at the GSI accelerator facilities in Darmstadt (Germany) using the SIS18 synchrotron combined with
the fragment separator FRS. The FRS is a zero-degree high-resolving power spectrometer with a typical resolving power
of $B\rho/\Delta B\rho =1500$ and with an angular acceptance
of $\pm$15 mrad around the central trajectory. In the present work, beams of $^{112}$Sn, accelerated up to kinetic energies of 1$A$ GeV, 
were delivered by the SIS18 synchrotron with an intensity around 10$^{8}$ ions per spill and 
were then guided up to the FRS entrance to produce the isobaric charge-exchange reactions. 
The FRS spectrometer was used in the achromatic mode, where
the reaction target is placed at the FRS entrance and the full spectrometer
is then utilized to separate and identify the nuclear residues, as shown in Fig.~\ref{fig:1}.

%%%%%%%%%%%%%%%%%%%%%%%%%%%%%%%%%%%%%%%
%%%%%%%%%%%%%%%%%%%%%%%%%%%%%%%%%%%%%%%

Two separate measurements were carried out using polyethylene (CH$_{2}$) and carbon targets with a thickness of 95 and 103 mg$/$cm$^{2}$, respectively.
The experimental data with the carbon target were also used to subtract the contribution of the carbon-induced reactions in the CH$_{2}$ target
and obtain the contribution of the reactions coming from the interactions with protons. An additional measurement was performed without targets
in order to obtain the background of reactions coming from other layers of matter located at the FRS target area, which was of around 1 $\%$.

Nuclei transmitted through the FRS can be identified in mass over charge ($A/q$) through the determination of
their magnetic ridigity ($B \rho$) and velocity ($v$) according to the following equation:
\begin{equation}
\frac{A}{q} = \frac{ e B \rho }{ u \gamma \beta c }
\end{equation}
where $q$ is the atomic number if we assume that the fragments were completely stripped ($q=Z$), $u$ is the atomic mass unit, 
$e$ is the elementary charge, $c$ is the speed of the light, $\beta=v/c$ and $\gamma=1/\sqrt{1-\beta^2}$.

The magnetic ridigity $B \rho$ of each reaction product can be obtained in terms of the magnetic-ridigity value
of an ion following the central trajectory along the FRS and its position obtained from the time-projection chambers (TPC), according
to the dispersive coordinate at the intermediate focal plane (F2), using the equation:
\begin{equation}\label{eq:planeS2}
 B \rho =B \rho_{0}^{F2} \left(1- \frac{x_{2} }{ D_{F2}} \right)
\end{equation}
where $x_{2}$ is the horizontal position at the intermediate focal plane determinated
with a resolution of 300 $\mu m$ full width at half maximum (FWHM)~\cite{Janik11}, $D_{F2}$ is the
value of the dispersion from the production target until the focal plane F2, and $B \rho_{0}^{F2}$ is
the magnetic ridigity of a central trajectory along the first section of the FRS. The magnetic ridigity of the central trajectory
as well as the dispersion were determinated by measuring the trajectory of $^{112}$Sn projectiles at 1$A$ GeV
for different values of the magnetic fields in the dipoles of the spectrometer. These measurements allowed us to calibrate the FRS optics,
obtaining a dispersion for the middle focal plane of $D_{F2}=-7.48\pm0.02$ cm/$ \% $.

\begin{center}
\begin{figure}[h]
\begin{center}
\includegraphics[width=0.45\textwidth,keepaspectratio]{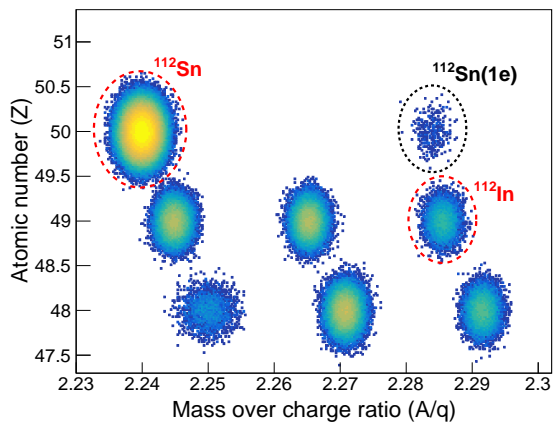}
\caption{(Color online) Identification matrix at the final focal plane of the FRS obtained from reactions induced
by $^{112}$Sn impinging on a carbon target. 
The figure was obtained by overlapping two magnetic settings of the FRS centered on the isotopes of $^{112}$Sn and $^{112}$In. 
The calibration in atomic and mass number was performed with respect to the signals registered for the beam of $^{112}$Sn. 
We also indicate the single electron charge state of $^{112}$Sn with a dotted ellipse. The figure shows the good resolution 
in atomic and mass number achieved in this measurement: $\Delta Z/Z$=6~$\times$~10$^{-3}$ 
and $\Delta A/A$=1~$\times$10$^{-3}$ (FWHM), respectively.
}
\label{fig:2}
\end{center}
\end{figure}
\end{center}

Using Eq.\ (\ref{eq:planeS2}) together with the velocity of fragments obtained from time-of-flight (ToF) measurements performed with the plastic scintillators
located at the focal planes of the FRS, we get the identification in 
mass-over-charge $(A/q)$ of each fragment applying the method described in Refs. \cite{Teresa14,JL2017}. 
In addition, the measurement of the atomic number $Z$ provided by ionization chambers located in the focal planes of
the FRS allows for the complete identification of the residual nuclei. 
As a result, Fig.\ \ref{fig:2} shows the identification matrix of the fragmentation residues measured from the projectiles of $^{112}$Sn
impinging on the carbon target.

Isobaric charge-exchange cross sections were accurately determined
by normalizing the production yield of the charge exchange residual nuclei to
the number of projectiles and target nuclei. The number of incoming projectiles was obtained by using
a secondary electron monitor (SEETRAM)~\cite{Junghans1996} placed at the entrance of the FRS,
which was calibrated with a reference plastic scintillator~\cite{Voss1995}.
The uncertainty in the determination of the incoming projectiles was around 3.5$\%$.
The number of charge exchange residues is obtained from the corresponding identification matrix at the
final focal plane by gating on the fragment of interest, as shown in Fig.~\ref{fig:2} for the isobaric charge-exchange residue of $^{112}$In.

The missing-energy spectra of the recoiling $(p, n)$ charge-exchange residues were obtained from the measured momentum of the residual
nuclei, which was also corrected by the dependence on the beam-extraction time given by the synchrotron SIS18~\cite{Tanaka18} 
to reach the best FRS accuracy.
Thanks to the high-resolving power of the FRS together with a sizeable reduction of matter along the beam line, we allowed to obtain
this spectrum with a resolution of around 10 MeV, which is a factor 2 better than that was obtained at SATURNE experiments~\cite{roy1988}.

%%%%%%%%%%%%%%%%%%%%%%%%%%%%%%%%%%%%%%%
%%%%%%%%%%%%%%%%%%%%%%%%%%%%%%%%%%%%%%%

\begin{center}
\begin{figure}[h]
\begin{center}
\includegraphics[width=0.45\textwidth,keepaspectratio]{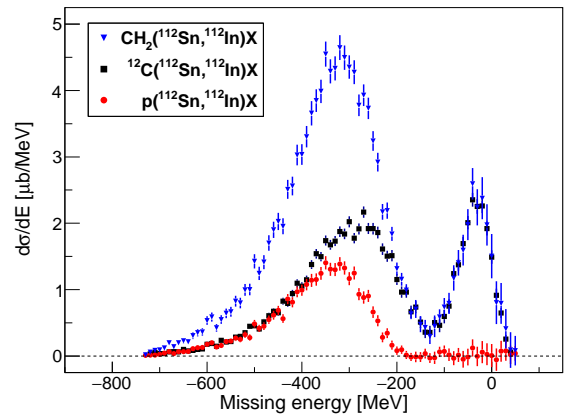}
\caption{(Color online) Missing-energy spectra obtained from the CH$_2$, $^{12}$C and proton targets for the isobaric charge-exchange reaction ($^{112}$Sn,$^{112}$In).
}
\label{fig:3}
\end{center}
\end{figure}
\end{center}

In Fig. \ref{fig:3} we show the missing-energy spectra obtained from the $(p,n)$ isobaric charge-exchange reactions
induced by ions of $^{112}$Sn in the CH$_{2}$ (triangles) and $^{12}$C (squares) targets at projectile kinetic energies of 1$A$ GeV,
which were normalized to the cross section of each reaction. For both spectra, the background was subtracted on a bin-by-bin basis.
The contribution of the proton target (circles) was then obtained from the CH$_{2}$ spectrum by subtracting the carbon contribution, according to the equation:
$\sigma_p = 0.5 \times [\sigma_{CH_2} - \sigma_C]$, applied on a bin-by-bin basis as well.
These spectra show that the spin-isospin response of nuclei is concentrated in two energy domains. 
The first group of excitations starting at missing energies close to zero is of quasi-elastic origin. 
This region is dominated by quasi-free charge-exchange processes between projectile and target nucleons, 
primarily reflecting global intrinsic properties of the colliding nuclei like Fermi motion and, to a lesser extent, 
isovector mean-field dynamics. Collective isovector multipole modes will contribute as well, but their identification 
would require a multipole decomposition of the spectra, as used for light ion reactions \cite{PhysRevC.86.014304}, which is beyond the feasibility of the present analysis. 
Note that  the quasi-elastic peak, clearly seen in the case of the spectra from CH$_{2}$ and C targets, disappears, as expected, when both spectra are combined to obtain
that from the proton target. At lower missing energies, the observed peak corresponds to a nucleon being excited into a $\Delta$ resonance in the target or projectile.
This twofold spectrum is a common feature of all isobaric charge-exchange reactions induced at high kinetic energies (more than 0.4$A$ GeV)
and illustrates clearly that the $\Delta$ is the first spin-isospin excited mode of the nucleon, corresponding to a $\Delta S=1$ and $\Delta T=1$ spin-isospin change.

%%%%%%%%%%%%%%%%%%%%%%%%%%%%%%%%%%%%%%%

%%%%%%%%%%%%%%%%%%%%%%%%%%%%%%%%%%%%%%%%%%%%%%%%%%%%%%%
\section{Model calculations}
\label{sec:model}

Before discussing the results, we briefly describe the model we have employed to analyze the missing-energy
spectra obtained in the experiment. The nuclear structure model was presented in part already in Refs. \cite{Lenske:2018bgr,Lenske:2018bvq}. Details of the reaction approach will be discussed in a forthcoming publication \cite{Vidana20}.

Under nuclear structure aspects, the reactions are probing in the quasi-elastic region nucleon hole-nucleon particle 
($N^{-1}N'$) configurations and in the deep inelastic region nucleon hole-resonance particle ($N^{-1}N^*$) configurations. 
The appropriate theoretical description of such excitations is discussed in detail in \cite{Lenske:2018bgr,Lenske:2018bvq}. 
A suitable approach for large scale studies of global properties of nuclear spectroscopy over the whole energy range of 
interest is response function theory. Earlier applications to light-ion induced nuclear reactions are found e.g. 
in \cite{BAKER:1997235,RAMSTROM:2004108}.

The essence of the response function approach is to solve directly the Dyson-equation for the nuclear polarization propagator
$\Pi_{\alpha\beta}(\omega,\mathbf{q})$  \cite{FW:1971}, describing the response of a many-body system to an external perturbation, exerted by one-body transition operators $\mathcal{O}_{\alpha,\beta}$ on the ground state $|A\ran$:
\be\label{eq:PP}
\Pi_{\alpha\beta}(\omega,\mathbf{q})=\lan A|\mathcal{O}^\dag_\beta\mathcal{G}_A(\omega,\mathbf{q})\mathcal{O}_\alpha|A\ran
\ee
The evolution of the nucleus is described by the many-body Green function $\mathcal{G}_A$. The interacting polarization propagator, Eq.~\eqref{eq:PP}, is given in terms of the non-interacting counterpart  $\Pi^0_{\alpha\beta}$ and the intranuclear residual interactions $\mathcal{V}_\lambda$ by the Dyson-equation
\be
\Pi_{\alpha\beta}=\Pi^0_{\alpha\beta}+\sum_\lambda\Pi^0_{\alpha\lambda}\mathcal{V}_\lambda\Pi_{\lambda\beta}
\ee
where $\Pi^0_{\alpha\beta}$ is defined according to Eq.\eqref{eq:PP} but with the Green-function $\mathcal{G}^0_A$ which does not include the residual interactions. In our case, we are interested in $N^{-1}N'$ and $N^{-1}N^*$ excitations. 
The extensions to that scenario are on the formal level easily incorporated.
For a single resonance, chosen here as $N^*=\Delta(1232)$, the polarization tensor is given by a 2-by-2 tensorial structure with $\Pi^{0}_{\alpha\beta}=diag\left( \Pi _{N\alpha,N\beta}^{(0)} ,\Pi _{\Delta\alpha,\Delta\beta }^{(0)}\right)$, interacting through and becoming mixed by
\be\label{eq:VND}
\mathcal{V}=
\left( {\begin{array}{*{20}{c}}
{{\mathcal{V}_{NN}}}&{{\mathcal{V}_{N\Delta }}}\\
{{\mathcal{V}_{\Delta N}}}&{{\mathcal{V}_{\Delta \Delta }}}
\end{array}} \right)
\,.
\ee
The indices indicate the particle-type components of the excitations, e.g. $N\alpha,N\beta$ stands for the propagation of the $N^{-1}N'\to N^{-1}N'$ . Each additional $N^{-1}N^*$ channel adds another block of interactions and polarization propagators. For the present application, the Dyson-equation is solved in infinite asymmetric nuclear matter. The response functions are obtained 
in the local density approximation by replacing proton and neutron matter densities $\rho_{p,n}$ by their radial dependent counterparts of the interacting nuclei, indicated below by $\rho_A(\mathbf{r})=\lan A|\hat{\rho}|A\ran$. The charged current nuclear response functions are defined by
\be
R^{(A)}_{B}(\omega,\mathbf{q})=
-\frac{1}{\pi}Im\left(\sum_\gamma\int d^3r \Pi_{B\gamma,B\gamma}(\omega,\mathbf{q}|\rho_A(\mathbf{r})) \right).
\ee
which describe the spectral distributions in a finite nucleus for the baryon configurations $B=N,N^*$, summed over interactions $\gamma$.
For a detailed discussion on the formalism and practical applications, we refer to our recent review articles~\cite{Lenske:2018bgr,Lenske:2019iwu,Lenske:2018jav}, 
where results for quasi-elastic heavy ion charge exchange reactions and $N^{-1}N$ and $N^{-1}\Delta$ excitations in ($e,e'$) scattering are found. 

\begin{center}
\begin{figure}[t!]
\begin{center}
\includegraphics[width=0.46\textwidth,keepaspectratio]{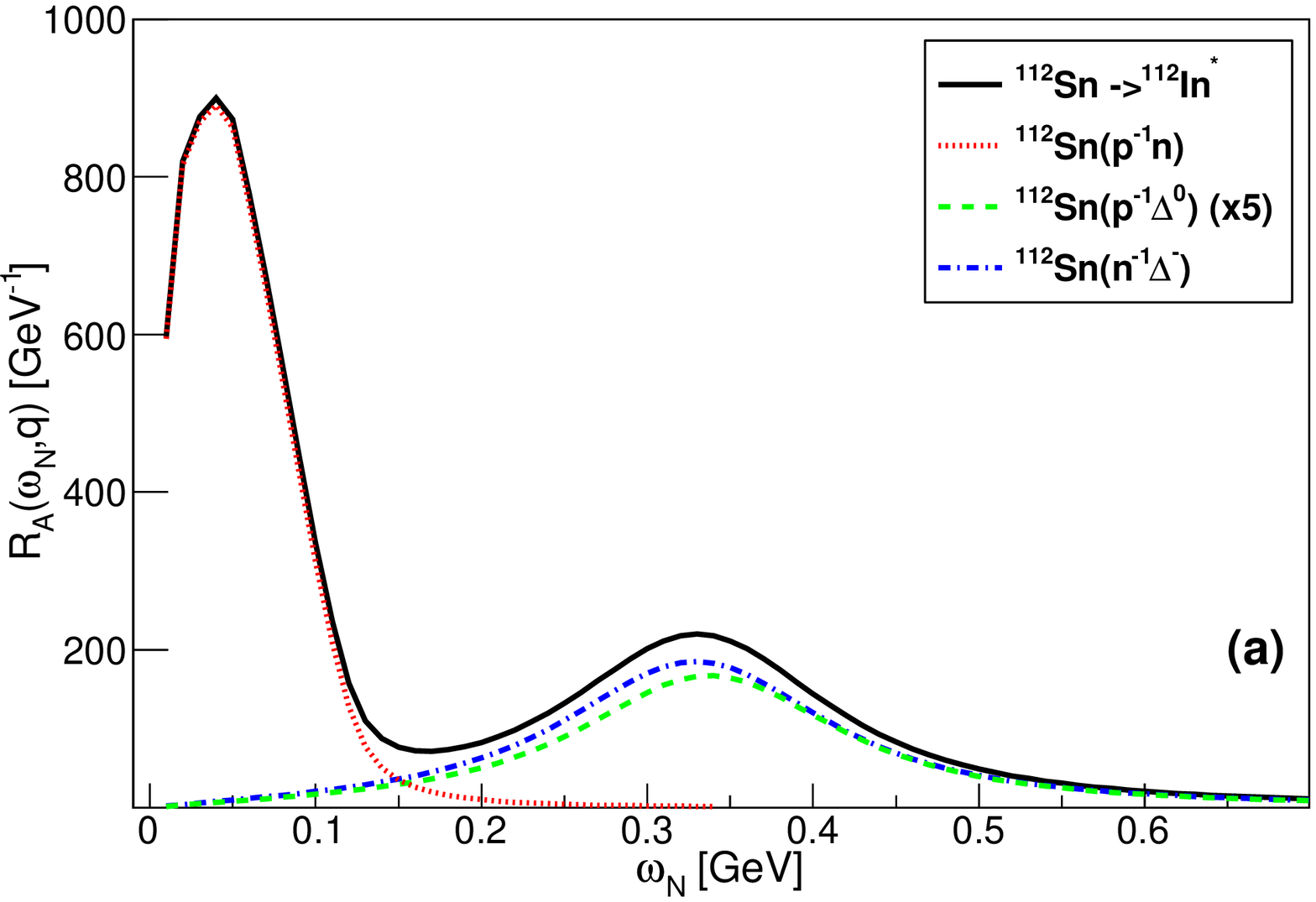}
\includegraphics[width=0.46\textwidth,keepaspectratio]{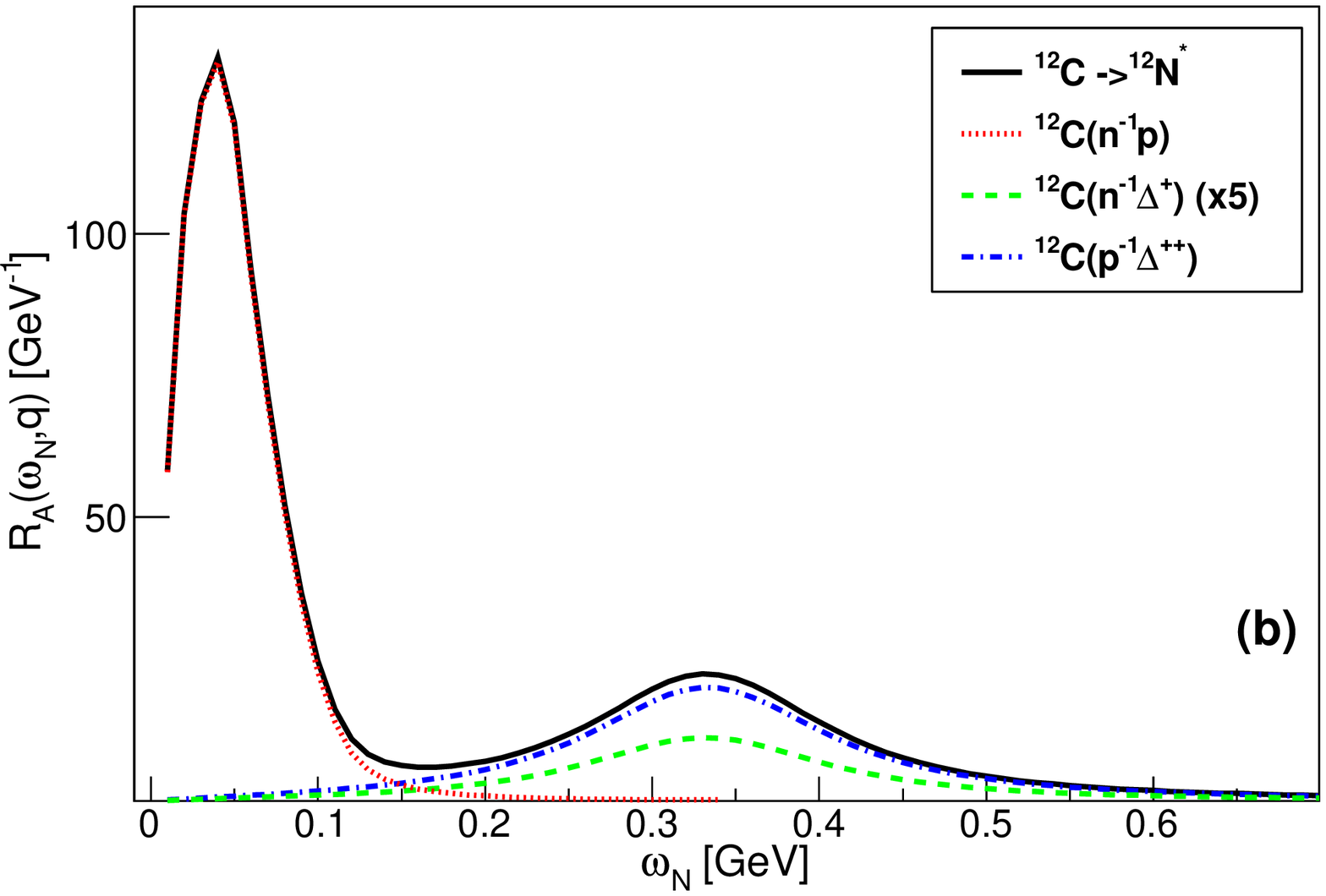}
\caption{(Color online) Response function for the transitions  $^{112}Sn\to {}^{112}In^*$ (a) 
and $^{12}C\to {}^{12}N^*$ (b) as a function of the excitation energy with respect to the nucleon at three-momentum transfer $q=300$~MeV/c. 
The quasi-elastic peak due to  $p^{-1}n$ and $n^{-1}p$ excitations, respectively, and the inelastic components from the various $N^{-1}\Delta$ 
excitations are displayed. The charged pion emission threshold opens at $\omega_N\sim 139$~MeV.
}
\label{fig:4}
\end{center}
\end{figure}
\end{center}

As typical results the response functions for $^{112}Sn\to {}^{112}In^*$ and $^{12}C\to {}^{12}N^*$ are shown in Fig.~\ref{fig:4} as a function of the 
excitation energy with respect to the nucleon. The response functions are evaluated for the transition density 
operator $\delta\hat{\rho}=\hat{\rho}-\rho_A$ at a fixed three-momentum transfer, being summed over the excitation channels $B$. 
The interactions $\mathcal{V}_\lambda$ are derived in Landau-Migdal theory from the GiEDF energy density 
functional \cite{Tsoneva:2017kaj,Lenske:2019ubp}, based on a G-matrix interaction supplemented by three-body forces. 
The calculations include in-medium self-energies for nucleons and the Delta states. 
For the latter, also the self-energies from the pionic decay channels are taken into account as discussed in Ref.~\cite{Lenske:2018bgr}.

\begin{center}
\begin{figure}[t]
\begin{center}
\includegraphics[width=0.32\textwidth,keepaspectratio]{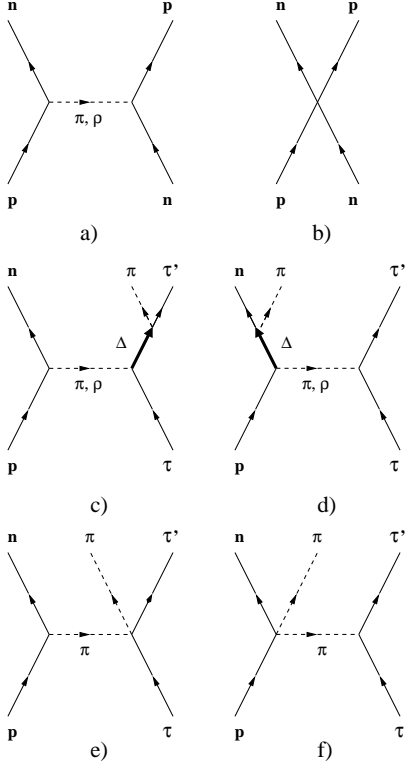}
\caption{
(Color online) Direct contributions to the quasi-elastic (diagrams a and b), and ineslastic (diagrams c to f) $(p,n)$ elementary processes.
The emitted pion in the final state is assumed to come either from the decay $\Delta\rightarrow N\pi$ (diagrams c and d) or to be produced 
in s-wave without the excitation of any resonance (diagrams e and f). 
Exchange contributions are obtained by interchanging the lines of the incoming proton of the projectile and the nucleon $\tau$ of the target.
}
\label{fig:diagram}
\end{center}
\end{figure}
\end{center}

The reaction model includes the direct and exchange contributions to the missing-energy spectrum from quasi-elastic as well as inelastic
$(p,n)$ elementary processes  which are described in terms of the exchange of virtual $\pi$- and $\rho$-mesons between the interacting nucleons.
The Feynman diagrams corresponding to the direct contributions to the elementary processes considered in the model are shown in Fig.~\ref{fig:diagram}.
The complete list of these processes is given in Table~\ref{tab:elp} for the two reactions studied,
p($^{112}$Sn,$^{112}$In)X and $^{12}$C($^{112}$Sn,$^{112}$In)X.
The exchange contributions can be simply obtained by interchanging the lines of the incoming proton of the projectile and the nucleon $\tau$ of the target. We note that the nucleon $\tau$ can only be a proton in the case of the p($^{112}$Sn,$^{112}$In) reaction but it can be either a neutron or a proton in the case of the $^{12}$C($^{112}$Sn,$^{112}$In)X one.
The excitation of the $\Delta$ resonance and its subsequent decay into a nucleon and a pion is considered both in the target (diagram c) and in the projectile (diagram d).
Note that the label $\Delta$ in diagrams c and d indicate schematically the isospin states $\Delta^{++},\Delta^+$ or $\Delta^0 $, depending on the particular elementary process
(see Table~\ref{tab:elp}). We note that contributions from the excitation of other resonances, such as for instance the Roper $N^{*}$, 
have been ignored in the model. Contributions where the emitted pion is produced in s-wave are, however, considered (diagrams e and f).
The basic ingredients of the model are the $NN\pi$, $NN\rho$ and $N\Delta\pi$ vertices \cite{Ericson87,Fernandez92} 
and the $\pi N \rightarrow \pi N$ s-wave amplitude~\cite{Oset89,Fernandez92}. 
The inclusive cross section for the reaction A(a,b)B describing the differential missing-energy spectrum of the outgoing 
projectile-like ion only is given by the sum of a quasi-elastic (qe) and an inelastic (in) contribution

\begin{equation}
 \frac{d\sigma}{dE_{b}} = \frac{d\sigma}{dE_{b}} \Big |_{qe}+ \frac{d\sigma}{dE_{b}} \Big |_{in} \ .
\end{equation}
The quasi-elastic contribution is obtained as
\begin{equation}
 \frac{d\sigma}{dE_{b}} \Big |_{qe}=\langle N_{pn}\rangle |\mathcal{M}_{qe}|^2
\label{eq:dxsqe}
\end{equation}
being $\langle N_{pn}\rangle$ the average number of projectile protons and target neutrons participating in the reaction defined through Eq.\ (\ref{eq:av}) and
\bea
&&\left|\mathcal{M}_{qe}\right|^2=\frac{2|{\vec p}_b|}{(2\pi)^2}\frac{m^4}{\lambda^{1/2}(s,m^2,m^2)}   \nonumber \\
&&\times
\int  \frac{d\vec q }{(2\pi)^3}\frac{d\Omega_{b}}{E_{B}}
R^{(A)}_N(\sqrt{s}-E_b,\vec{q})\langle \left|M_{qe}(\vec {q})\right|^2\rangle \nonumber \\
\label{eq:dxsqe}
\eea
where $s$ is the total energy in the center-of-mass frame, $m$ is the nucleon mass, $\langle \cdot \rangle$ indicates the average and sum over the initial and final spins, $R^{(A)}_N(\omega,\mathbf{q})$
is the spectral distribution  of nucleonic $N'N^{-1}$ target transitions, $M_{qe}$ is the scattering amplitude of the elementary process $p+n\rightarrow n+p$ and
$\lambda(a,b,c)=a^2+b^2+c^2-2ab-2ac-2bc$ is the so-called K\"{a}llen function.

The inelastic contribution is given by
\begin{eqnarray}
\frac{d\sigma}{dE_{b}} \Big |_{in}&=&\langle N_{pp}\rangle |\mathcal{M}_{in}^{(pp\rightarrow np\pi^+)}|^2 \nonumber \\
&+&\langle N_{pn}\rangle |\mathcal{M}_{in}^{(pn\rightarrow nn\pi^+)}|^2 \nonumber \\
&+&\langle N_{pn}\rangle |\mathcal{M}_{in}^{(pn\rightarrow np\pi^0)}|^2
\end{eqnarray}
with the average number of projectile and target protons $\langle N_{pp}\rangle$ participating defined also through Eq.~(\ref{eq:av}) and
\bea
&&\left|\mathcal{M}_{in}^{(p\tau\rightarrow n\tau'\pi)}\right|^2= \frac{1}{S}\frac{|{\vec p}_b|}{(2\pi)^5}\frac{m^4}{\lambda^{1/2}(s,m^2,m^2)}  \nonumber \\
&&\times
\int\frac{d\vec q }{(2\pi)^3} \frac{d\vec p_\pi d\Omega_{b}}{E_{B}E_\pi}   \nonumber \\
&&\times R^{(A)}_{\Delta}(\sqrt{s}-E_b-E_\pi,\vec{q}-\vec{p}_\pi) \nonumber \\
&&\times\langle \left|M_{in}^{ (p\tau\rightarrow n\tau'\pi )}(\vec{q}-\vec{p}_\pi) \right|^2\rangle \ . \nonumber \\
\label{eq:dxsin}
\eea

Here $M_{in}^{(p\tau\rightarrow n\tau'\pi)}$ is the scattering amplitude of the elementary process $p+\tau \rightarrow n+\tau'+\pi$ and $S$ 
is the symmetry factor,
\begin{equation}
S=\prod_{l}k_l!
\label{eq:symfact}
\end{equation}
for $k_l$ identical particles of species $l$ in the final state. 
In our case $S$ can be 1 or 2 depending on the particular reaction channel (see Table~\ref{tab:elp}).

%%%%%%%%%%%%%%%%%%%%%%%%%%%%%%%%%%%%%%%
%%%%%%%%%%%%%%%%%%%%%%%%%%%%%%%%%%%%%%%

\begin{table}[t!]
\caption{Elementary processes included in the description of the isobaric charge-exchange reactions p($^{112}$Sn,$^{112}$In)X 
and $^{12}$C($^{112}$Sn,$^{112}$In)X.
}
\label{tab:elp}
\begin{center}
\small
\begin{tabular}{cc}
\hline
\hline
\\
\multicolumn{2}{c}{p($^{112}$Sn,$^{112}$In)X } \\ \\ \hline
\multicolumn{1}{c}{inelastic (target)} &  \multicolumn{1}{c}{inelastic (projectile)} \\ \hline
\\
$p(p,n)\Delta^{++}\rightarrow p(p,n)p\pi^+$  & $p(p,\Delta^+)p\rightarrow p(p,n\pi^+)p$ \\
$p(p,n)p\pi^+$ (s-wave) & $p(p,n\pi^+)p$ (s-wave)  \\
\\
\hline
\hline
\\
\multicolumn{2}{c}{$^{12}$C($^{112}$Sn,$^{112}$In)X} \\ \\
\hline
\\
quasi-elastic &{$n(p,n)p$}
\\
\\
\hline
\\
\multicolumn{1}{c}{inelastic (target)} &  \multicolumn{1}{c}{inelastic (projectile)} \\
\\
\hline
\\
$p(p,n)\Delta^{++}\rightarrow p(p,n)p\pi^+$  & $p(p,\Delta^+)p\rightarrow p(p,n\pi^+)p$ \\
$p(p,n)p\pi^+$ (s-wave) & $p(p,n\pi^+)p$ (s-wave)  \\ \\
$n(p,n)\Delta^+\rightarrow n(p,n)n\pi^+$  & $n(p,\Delta^+)n\rightarrow n(p,n\pi^+)n$ \\
$n(p,n)\Delta^+\rightarrow n(p,n)p\pi^0$  & $n(p,\Delta^0)n\rightarrow n(p,n\pi^0)p$ \\
$n(p,n)n\pi^+$ (s-wave) & $n(p,n\pi^+)n$ (s-wave) \\
$n(p,n)p\pi^0$ (s-wave)  & $n(p,n\pi^0)p$ (s-wave) \\
\\
\hline
\hline

\end{tabular}
\end{center}
\end{table}
%%%%%%%%%%%%%%%%%%%

The average number of proton-neutron (pn) and proton-proton (pp) elementary processes contributing to the reaction is calculated as
\begin{equation}
\langle N_{p\tau}\rangle=\langle N_{p}^{(P)}\rangle\times\langle N_{\tau}^{(T)}\rangle \ , \,\,\,\, \tau=n,p
\label{eq:av}
\end{equation}
where $\langle N_{p}^{(P)}\rangle$ and $\langle N_{\tau}^{(T)}\rangle$ are, respectively, the average number of protons 
in the projectile and neutrons or protons in the target participating in the reaction. These numbers are calculated as
\begin{eqnarray}
\langle N_{p}^{(P)}\rangle&=&Z_P\frac{\sigma_T}{\sigma_{PT}}    \\
\langle N_{n}^{(T)}\rangle&=&(A_T-Z_T)\frac{\sigma_P}{\sigma_{PT}}  \\
\langle N_{p}^{(T)}\rangle&=&Z_T\frac{\sigma_P}{\sigma_{PT}}  \ ,
\end{eqnarray}
where $A_P$ $(Z_P)$  and $A_T$ $(Z_T)$ are the mass (atomic) numbers of the projectile and target nuclei, 
respectively and  $\sigma_{P(T)}$ and $\sigma_{PT}$ the total nucleon-nucleus and nucleus-nucleus cross section
determined by using the Glauber model
\begin{equation}
\sigma_{P(T)}=\int d\vec b \Big(1-(1-T_{P(T)}(b)\sigma_{NN})^{A_{P(T)}} \Big)
\end{equation}
\begin{equation}
\sigma_{PT}=\int d\vec b \Big(1-(1-T_{PT}(b)\sigma_{NN})^{A_{P}A_{T}} \Big)
\end{equation}
with $\sigma_{NN}$ the nucleon-nucleon cross section, for which we take here a value of 40 mb, and
\begin{equation}
T_{P(T)}(b)=\int dz \rho_{P(T)}(b,z)
\end{equation}
\begin{equation}
T_{PT}(b)=\int d\vec s T_P(|\vec s -\vec b|)T_T(s) \ ,
\end{equation}
with the projectile and target densities $\rho_P$ and $\rho_T$ obtained from a relativistic mean field model calculation using the FSU model \cite{Todd05}.

%%%%%%%%%%%%%%%%%%%%%%%%%%%%%%%%%%%%%%%%%%%%%%%%%%%%%%%
\section{Results and discussion}
\label{sec:res}

The spectra shown in Fig.~\ref{fig:3} are affected by the energy and angular straggling of ions passing
through the targets located at the FRS entrance, as observed in previous experiments \cite{Kelic04,vargas13}. In order to subtract these contributions an unfolding procedure,
based on the Richardson-Lucy's technique~\cite{Lucy74} together with a regularization
method to optimize the stability of the solution against statistical fluctuations~\cite{vargas13}, was used to improve the sharpness of the spectrum.
The unfolding was carried out using as FRS response function the missing-energy spectrum of the primary beam of $^{112}$Sn passing
through the corresponding target after subtracting the local energy-straggling due to the exchange of a proton with the target nuclei. 
This leads to the response function of an ion of $^{112}$In.

%%%%%%%%%%%%%%%%%%%%%%%%%%%%%%%%%%%%%%%
\begin{center}
\begin{figure}[h!]
\begin{center}
\includegraphics[width=0.45\textwidth,keepaspectratio]{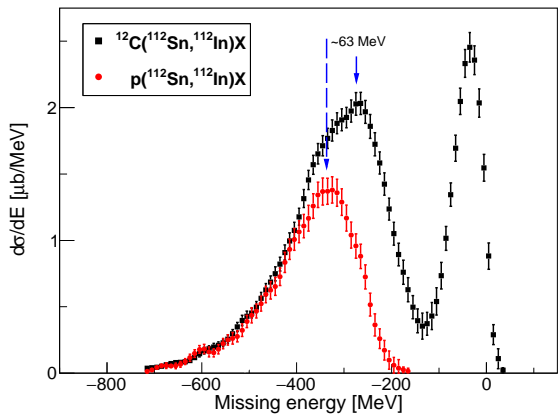}
\caption{(Color online) Missing-energy spectrum for the $(p,n)$ isobaric charge-exchange reactions 
induced by ions of $^{112}$Sn in carbon and proton targets.
}
\label{fig:6}
\end{center}
\end{figure}
\end{center}
%%%%%%%%%%%%%%%%%%%%%%%%%%%%%%%%%%%%%%%

\begin{center}
\begin{figure}[h!]
\begin{center}
\includegraphics[width=0.45\textwidth,keepaspectratio]{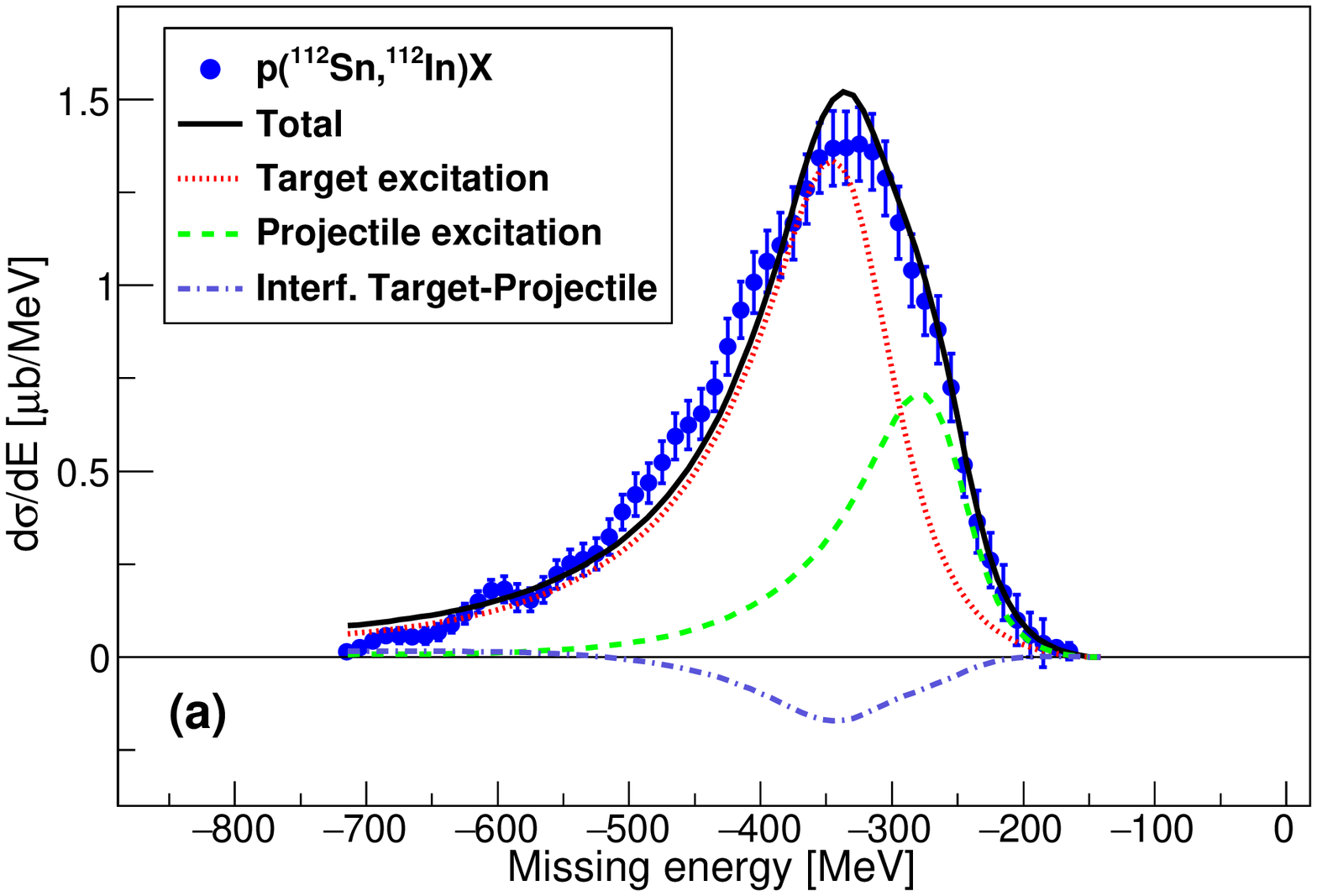}
\includegraphics[width=0.45\textwidth,keepaspectratio]{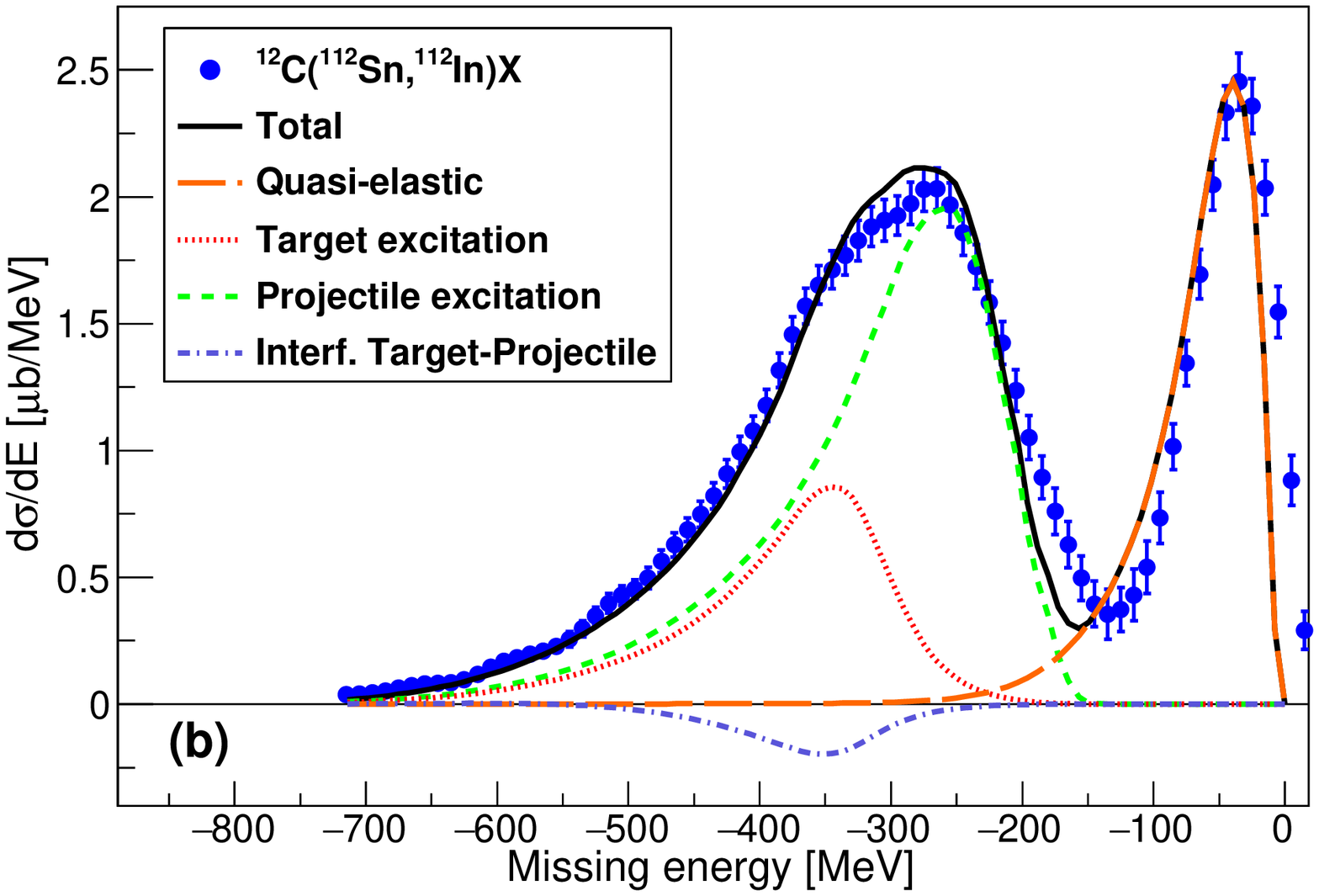}
\caption{(Color online) Comparison of the measured missing-energy spectra for the proton (a) and carbon (b) targets with the model calculations. 
The separate contributions from the quasi-elastic and inelastic (target and projectile excitation) processes are shown. 
The interference term between the target and projectile amplitudes is also shown.
}
\label{fig:7}
\end{center}
\end{figure}
\end{center}

The results of this unfolding procedure are displayed in Fig.~\ref{fig:6} for the carbon (squares) and proton (circles) targets.
As it is seen in the figure, at first glance, there seems to be a shift of about 63~MeV in the position of the $\Delta$ peak between both measurements. 
This observation is consistent with the results obtained from the SATURNE experiments~\cite{cotardo86}.
However, by looking with more detail the missing-energy spectrum obtained  with the carbon target, one can observe, at the left of the peak, 
the presence of a little shoulder positioned essentially at the same energy of the peak observed in the measurement 
with the proton target. In order to understand better this feature, in Fig.~\ref{fig:7} we compare our experimental data
for both measurements with the model calculations described in Sec.\ \ref{sec:model}. The shape of the experimental spectrum of both reactions is 
reasonably well reproduced by the model. The separate contributions from the quasi-elastic and inelastic (target and projectile excitation) processes 
are shown, as well as the interference between the target and projectile excitation processes. 
We should note that in the calculation for the $^{12}$C($^{112}$Sn,$^{112}$In)X reaction these contributions have been rescaled separately for comparison with the data. 
We will see in the following that, in addition, 
the model allows to reveal a large amount of information on the different reaction mechanism that lead to the inclusive spectra measured.

Let us analyze first the spectrum obtained with the proton target. The first thing we observe is, as expected in this case, 
the absence of the quasi-elastic peak at missing-energies close to zero. At lower missing energies the spectrum can be explained 
as a result of the superposition of the four elementary processes (see Table~\ref{tab:elp}) contributing to the reaction p($^{112}$Sn,$^{112}$In)X.
The spectrum is dominated by the excitation of the $\Delta^{++}$ in the target, although the contribution of the other three elementary processes, 
particularly the $\Delta^+$ excitation in the projectile, is necessary to reproduce its shape. See in particular the small shoulder observed 
at the right of the peak that can be well understood mainly thanks to the contribution of the $\Delta^+$ excitation in the projectile. 
The contribution from the two processes where the pion emitted in s-wave is small compared to those of the $\Delta^{++}$ and $\Delta^+$ excitations. 
Note that the shape and the position of the resonance is different if it is excited in the target or in the projectile. 
This is because its invariant mass is different in both cases. When the $\Delta^{++}$ is excited in the target, 
its invariant mass does not depend on the momentum of the emitted pion and, therefore, the scattering amplitude 
can be taken out of the integral in Eq.~(\ref{eq:dxsin}). In this case, the shape of the resonance appears almost 
symmetric and it has a peak at approximately -345~MeV. On the contrary, when the $\Delta^+$ 
is excited in the projectile, its invariant mass depends explicitly on the momentum of the emitted pion and, therefore, 
in this case the scattering amplitude should also be integrated. Consequently, the shape of the resonance becomes more 
asymmetric and its position is shifted to the right by $\sim$ 63 MeV. Note that this is precisely the shift in the 
position of the $\Delta$ peak that seems to be observed between the measurements with the proton and carbon target.

Finally, let us analyze the results of the $^{12}$C($^{112}$Sn,$^{112}$In)X reaction. 
We observe that the quasi-elastic peak is reasonably well described by the model and that
the low missing-energy region of the spectrum is once more the result of the superposition of 
elementary processes where the excitation ($\Delta$ resonance or pion production in s-wave)
occurs either in the target or in the projectile. Note that in this case there are eight 
additional elementary processes (see Table\ \ref{tab:elp}) contributing
that were absent in the case of the reaction with the proton target, explaining the larger magnitude of the spectrum in this case, 
apart from the trivial scaling by the increased number of participating target nucleons. In addition, 
there is a change of the relative magnitude between the excitation in the target and the projectile. 
Whereas in the proton target case the dominant contribution is the excitation of the $\Delta^{++}$ in the target, 
now is that of the $\Delta^+$ in the projectile.
The reason is simply that in a complex target nucleus only part of the energy transferred goes into the excitation of 
the $\Delta^{++}$ resonance or the direct s-wave production of the pion. 
Besides the excitations of $^{12}$C($n^{-1}\Delta^+$) configurations (see Fig.~\ref{fig:4}), 
another part of the energy loss is employed in reactions such as, for instance, the knockout of nucleons from the target or its fragmentation. 
Consequently, this contribution is reduced with respect to that of the projectile excitation and is now observed 
simply as a shoulder at $\sim$ 63 MeV to the left of the $\Delta^{+}$ peak which, in this case, is the dominant one. 
This seems to indicate that the apparent shift in the position of the $\Delta$-resonance peak, observed in targets heavier than the proton, 
can be simply interpreted as a change in the relative magnitude between the contribution of the excitation of the resonance in 
the target and in the projectile. A similar conclusion was already pointed out by 
Oset, Shiino and Toki \cite{Oset89} in their analysis of the ($^3$He,t) reaction on proton, deuteron and carbon targets.
We finish this section by emphasizing that this is the first time that the excitation of the $\Delta$ resonance in 
isobar charge-exchange reactions with heavy nuclei is studied both experimentally and theoretically.

%%%%%%%%%%%%%%%%%%%%%%%%%%%%%%%%%%%%%%%%%%%%%%%%%%%%%%%
\section{Summary and Conclusions}
\label{sec:sum}

Isobaric charge-exchange reactions induced by ions of $^{112}$Sn at energies of $1A$ GeV on different targets have been investigated at GSI using the FRS spectrometer.
The experimental setup allows us to measure the cross sections of these processes with high accuracy and to determine in coincidence
the missing-energy spectra of the corresponding ejectiles. Thanks to the high-resolving power of the magnetic spectrometer FRS we can clearly identify
in the missing-energy spectra the quasi-elastic and inelastic components corresponding to the nuclear spin-isospin response of nucleon-hole
and $\Delta$ resonance excitations, respectively.

We observe an apparent shift of the $\Delta$-resonance peak of about (63$\pm$5) MeV when comparing the missing-energy 
spectra obtained from the measurements with proton and carbon targets. 
This observation is consistent with the results obtained from the SATURNE experiments \cite{cotardo86}. 
However, a detailed analysis with a theoretical model for the reaction indicates that this observation can be simply interpreted as a
change in the relative magnitude between the contribution of the excitation of the resonance in the target and in the projectile. 
Future exclusive measurements tagging the pion emitted from the decay of the $\Delta$ resonance will allow to distinguish between target and projectile 
excitations unambiguously, and the option to study also other resonances like the different modes of the Roper $N^*$, 
providing relevant information to understand in much more detail 
the formation and evolution of nucleonic resonances in the nuclear medium. An especially exciting prospect for future work on 
the upcoming Super-FRS experiments at FAIR using the WASA calorimeter~\cite{SFRSExp,YT2020} 
is the ability to extend the investigations also to beams of $\beta$-unstable exotic nuclei, 
giving access for the first time to study resonance dynamics in manifestly asymmetric nuclear matter.

\section*{Acknowledgments}
The authors are grateful to the GSI accelerator staff for providing an intense and stable beam of $^{112}$Sn. 
We thank to L.~\'{A}lvarez-Ruso for reading the manuscript carefully. 
We also thank the support from Unidad de Excelencia Maria de Maeztu. %under project MdM-2016-0692-17-2. 
J.L.R.S. thanks the support from the Department of Education, Culture and University Organization of
the Regional Government of Galicia under the program of postdoctoral fellowships (\textit{ED481B-2017$/$002}). 
I.V. thanks the support from the \textit{COST Action CA16214}.

\section*{References}

\bibliography{mybibfile}

\end{document}